\documentclass[final, numberedheadings]{aipproc}

\layoutstyle{6x9}

\usepackage{bm}
\usepackage{slashed}
\usepackage{graphicx}
\usepackage{subfigure}
\makeatletter
\newcommand\setcaptype[1]{\def\@captype{#1}}
\makeatother

\usepackage[svgnames]{xcolor}
\usepackage[normalem]{ulem}

\begin{document}

\begin{flushright} ADP-10-7/T703 \end{flushright}

\title{Kaon fragmentation function from NJL-jet model}


\author{Hrayr H.~Matevosyan}{
  address={CSSM, School of Chemistry and Physics, \\
University of Adelaide, Adelaide SA 5005, Australia}
}

\author{Anthony W. Thomas}{
  address={CSSM, School of Chemistry and Physics, \\
University of Adelaide, Adelaide SA 5005, Australia}
}

\author{Wolfgang Bentz}{
  address={Department of Physics, School of Science,\\  Tokai University, Hiratsuka-shi, Kanagawa 259-1292, Japan} 
}

\begin{abstract}
The NJL-jet model provides a sound framework for calculating the fragmentation functions in an effective chiral quark theory, where the  momentum and isospin sum rules are satisfied without the introduction of ad hoc parameters \cite{Ito:2009zc}. Earlier studies of the pion fragmentation functions using the Nambu--Jona-Lasinio (NJL) model within this framework showed good qualitative agreement with the empirical parameterizations. Here we extend the NJL-jet model by including the strange quark. The corrections to the pion fragmentation function and corresponding kaon fragmentation functions are calculated using the elementary quark to quark-meson fragmentation functions from NJL. The results for the kaon fragmentation function exhibit a qualitative agreement with the empirical parameterizations, while the unfavored strange quark fragmentation to pions is shown to be of the same order of magnitude as the unfavored light quark's. The results of these studies are expected to provide important guidance for the analysis of a large variety of semi-inclusive data.
\end{abstract}

\maketitle


\section{Quark Fragmentation and NJL-Jet Model}
\label{NJL-JET} 

 Quark fragmentation functions have long been of interest for analyzing hard scattering reactions  \cite{Field:1977fa,Altarelli:1979kv,Collins:1981uw,Jaffe:1996zw,Ellis:1991qj,Barone:2001sp,Martin:2003sk}. New experimental  efforts for extraction of fragmentation functions from deep-inelastic lepton-nucleon, proton-proton scattering and $e^{+}e^{-}$ annihilation data \cite{Hirai:2007cx,deFlorian:2007aj} have generated a renewed interest in the subject. The analysis of the transversity quark distribution functions \cite{Barone:2001sp,Ralston:1979ys} and a variety of other semi-inclusive processes \cite{Sivers:1989cc,Boer:2003cm} also critically depend on the knowledge of the fragmentation functions.
  
   The NJL-jet model of Ref. \cite{Ito:2009zc} provides a self-consistent framework for calculations of both quark distributions and fragmentation functions in an effective chiral quark theory. The advantage of the model is that  there are no ad hoc parameters introduced to describe fragmentation functions; the quark self-energy normalization factor in the coupled integral equations for the fragmentation functions arises naturally from the product ansatz. In this work we extend the NJL-jet model by introducing the strange quark, thus allowing fragmentation to both pions and kaons. The inclusion of the new channel is shown to bring significant corrections to the previously calculated fragmentation functions to pions.
   
\section{Including the Strange Quark}
\label{strange}

\subsection{Strange Quark Constituent Mass and Coupling}

Introducing the strange quark in the NJL-jet model involves calculating the strange quark distribution and fragmentation functions in the NJL framework, requiring the knowledge of the quark-meson (strange-light to kaon) coupling constant and the strange quark constituent mass. We calculate the quark-meson coupling using the same approach as Ref. \cite{Bentz:1999gx}. The strange quark mass is chosen semi-empirically to best fit the model calculated kaon decay constant to the experimental value.

 The quark-meson coupling constant is determined from the pole in the quark-antiquark t-matrix at the considered meson's mass. This involves the derivative of the familiar quark-bubble graph:

\begin{equation}
\label{EQ_VACUUM_BUBBLE}
\Pi(k)=2N_{c}i\int \frac{d^{4}q}{(2\pi)^{4}} Tr[\gamma_{5}S_{1}(q)\gamma_{5}S_{2}(q-k)],
\end{equation}

\begin{equation}
\label{EQ_COUPLING_KQQ}
\frac{1}{g^{2}_{mqq}}=- \left( \frac{\partial \Pi(k)}{\partial k^{2}}  \right)_{k^{2}=m^{2}_{m}}.
\end{equation}
 Evaluating  the integral over $q_{+}$ using the complex residue theorem and making the variable substitution $q_{-} = xk_{-}$ yields for the quark-meson coupling: 
 
\begin{eqnarray}
\label{EQ_MQQ_COUPL}
\frac{1}{g_{mqq}^{2}}& = & 2N_{c} \int_{0}^{1} dx \int \frac{ d^{2}q_{\perp}}{(2\pi)^{3}} \frac{ q_{\perp}^{2} + ((1-x)M_{1}+xM_{2})^{2} } {(q_{\perp}^{2} + (1-x)M_{1}^{2} +xM_{2}^{2} -x(1-x)m_{m}^{2} -i\epsilon)^{2} }.
\end{eqnarray}
 
 In this work we use the  Lepage-Brodsky (LB) ``invariant mass'' cut-off regularization as described in Refs \cite{Bentz:1999gx,Ito:2009zc}.  Using the experimental value of $f_{K}=0.114~{\rm GeV}$ yields a strange constituent quark mass $M_{s}=0.45~{\rm GeV}$ and the corresponding quark-kaon coupling constant of $g_{Kqq}=3$.

 \subsection{Quark Distribution and Fragmentation Functions}

The quark distribution function $f_{q}^{h}(x)$ has an interpretation as the probability to find a quark of type $q$ with momentum fraction $x$ in the hadron (in our case meson) $h$.  The corresponding cut diagram is shown in Fig. \ref{PLOT_Q_DISTR}a, which can be equivalently represented by the Feynman diagram depicted in Fig. \ref{PLOT_Q_DISTR}b:

\begin{figure}[ptb]
\centering 
\includegraphics[width=0.9\textwidth]{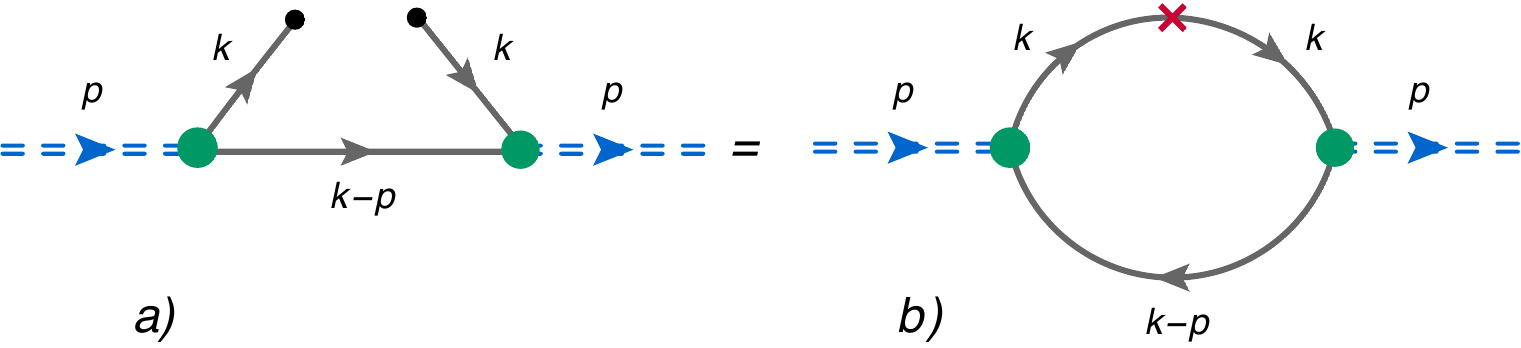}
\caption{Quark distribution functions.}
\label{PLOT_Q_DISTR}
\end{figure}

\begin{eqnarray}
\label{EQ_Q_DISTR}
f_{q}^{m}(x)& = &i N_{c} \frac{C_{I}}{2}   g_{mqq}^{2} \int \frac{ dk_{+} d^{2}k_{\perp}}{(2\pi)^{4}} Tr[\gamma_{5}S_{1}(k)\gamma_{+}S_{1}(k)\gamma_{5}S_{2}(k-p)] \\ \nonumber
&=&N_{c} C_{I} g_{mqq}^{2} \int \frac{ d^{2}k_{T}}{(2\pi)^{3}} \frac{k_{T}^{2}+((1-x)M_{1}+xM_{2})^{2}} {(k_{T}^{2}+(1-x)M_{1}^{2}+xM_{2}^{2}-x(1-x)m_{m}^{2})^{2}},
\end{eqnarray}
where $k_{-}= xp_{-}$ and $C_{I}$ is the corresponding flavor factor ($C_{I}=2$ for all the mesons considered except for $\pi_{0}$, where  $C_{I}=1$).

\begin{figure}[ptb]
\centering 
\includegraphics[width=0.45\textwidth]{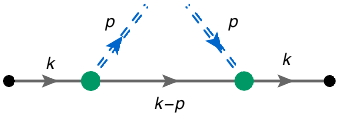}
\caption{Quark fragmentation functions.}
\label{PLOT_FRAG_QUARK}
\end{figure}

The elementary fragmentation function depicted in Fig. \ref{PLOT_FRAG_QUARK} can be written as:

\begin{eqnarray}
\label{EQ_QUARK_FRAG}
\nonumber
d_{q}^{m}(z)&=&N_{c} \frac{C_{I}}{2}  g_{mqq}^{2} \frac{z}{2} \int \frac{d^{4}k}{(2\pi)^{4}} Tr[S_{1}(k)\gamma_{+}S_{1}(k)\gamma_{5} (\slashed{k}-\slashed{p}+M_{2}) \gamma_{5}]\\ 
&& \times \delta(k_{-} - p_{-}/z) \delta( (p-k)^{2} -M_{2}^{2} )
=\frac{z}{2N_{c}} f_{q}^{m}(x = 1/z)\\ 
&=& \frac{C_{I}}{2}  g_{mqq}^{2} z \int \frac{ d^{2}p_{\perp}}{(2\pi)^{3}} \frac{p_{\perp}^{2}+((z-1)M_{1}+M_{2})^{2}} {(p_{\perp}^{2}+z(z-1)M_{1}^{2}+zM_{2}^{2}+(1-z)m_{m}^{2})^{2}}.
\end{eqnarray}

\section{Generalized NJL-Jet}
\label{SEC_NJL_JET}

The NJL-Jet model of Ref. \cite{Ito:2009zc} uses a multiplicative ansatz for the total fragmentation function to derive an integral equation for the quark cascade for the process depicted in Fig. \ref{PLOT_QUARK_CASCADE}. The derived integral equation for the total fragmentation function is:

\begin{figure}[ptb]
\centering 
\includegraphics[width=0.65\textwidth]{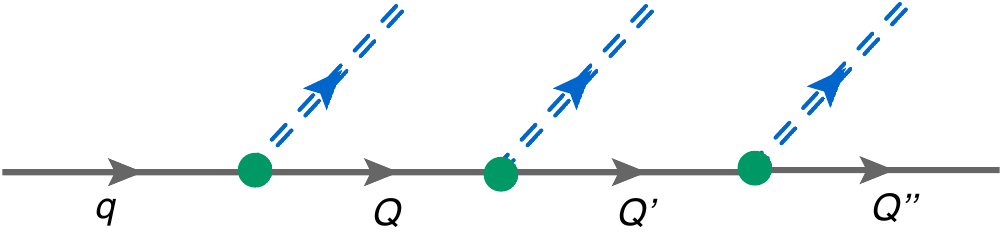}
\caption{Quark cascade.}
\label{PLOT_QUARK_CASCADE}
\end{figure}

\begin{eqnarray}
\label{EQ_JET_INT}
D^{m}_{q}(z)=\hat{d}^{m}_{q}(z)+\sum_{Q}\int^{1}_{z}\frac{dy}{y}\hat{d}^{Q}_{q}(\frac{z}{y})\; D^{m}_{Q}(y), \hspace{20pt}
\hat{d}^{Q}_{q}(z)=\hat{d}^{m}_{q}(1-z)|_{m=q\bar{Q}} .
\end{eqnarray}

Here $\hat{d}_{q}^{m}(z) = d_{q}^{m}(z)/(1-Z_{Q})$, where $Z_{Q}$ is the residue of the quark propagator in the presence of meson cloud (see Ref. \cite{Ito:2009zc}).  Then $\sum_{m}\int \hat{d}_{q}^{m}(z) dz=1$, thus allowing an interpretation as the probability of an elementary process. In Eq. (\ref{EQ_JET_INT}) the sum is over the flavor of the emitted quark $Q$ and the splitting function of quark $q$ into $Q$ with a momentum fraction $z$ is naturally the same as the splitting function of $q$ into a meson $m$ of a flavor composition $q\bar{Q}$ with a momentum fraction $1-z$.

 The result in Eq.(\ref{EQ_JET_INT}) resembles the integral equation ansatz of Field and Feynman's quark-jet model \cite{Field:1976ve,Field:1977fa}. Rewriting the above expression helps to elucidate the probabilistic interpretation of the model:

\begin{equation}
\label{ EQ_FRAG_PROB}
D^{m}_{q}(z)dz=\hat{d}^{m}_{q}(z)dz+\sum_{Q}\int^{1}_{z}\hat{d}^{Q}_{q}(y) dy  \; D^{m}_{Q}(\frac{z}{y}) \frac{dz}{y}.
\end{equation}

Here the left hand side term has the meaning of the probability to create a meson $m$ carrying the momentum fraction $z$ to $z+dz$ of initial quark $q$. The first term on the right hand side corresponds to the probability of creating the meson with momentum fraction $z$ to $z + dz$ in the first step of the cascade, plus the second term corresponding to the creation of the meson further down the quark cascade after a splitting to a quark $Q$ with momentum fraction $y$. Here the probability of creating the meson $m$ scales with the momentum fraction left to the quark after the splitting $z/y$, which is clearly only the case in the Bjorken limit. Thus the model can be trivially generalized by including the strange quark directly in the Eq. (\ref{EQ_JET_INT}). 

 We solve the coupled set of integral equations  using the elementary fragmentation functions of Eq. (\ref{EQ_JET_INT}) for $u$, $d$ and $s$ quark fragmentation to a given meson. The corresponding fragmentation functions for the anti-quarks are obtained using the charge symmetry from fragmentation functions of quarks to the corresponding anti-meson. The comparisons with the phenomenological parametrizations of Ref \cite{Hirai:2007cx} are performed by DGLAP evolving the calculated fragmentation functions from the low-energy model scale of $Q^{2}=0.18~{\rm GeV}^{2}$ to $4~{\rm GeV}^{2}$ at leading order using the software from Ref. \cite{Miyama:1995bd}. The evolution kernel for the distribution functions is modified as described in Appendix B of Ref. \cite{Ito:2009zc}

 It is easy to see, using the properties of $d_{q}^{m}$ along  with the normalization condition of $\hat{d}_{q}^{m}$, that the solutions of the integral equations (\ref{EQ_JET_INT}) should satisfy both momentum and isospin sum rules. Our numerical solutions obey these rules within  numerical errors of less than a percent.

\section{Results and Conclusions}
\label{results}

\begin{figure}[ptb]
\setcaptype{figure} 
\centering 
\subfigure[] {
\includegraphics[width=0.4\textwidth]{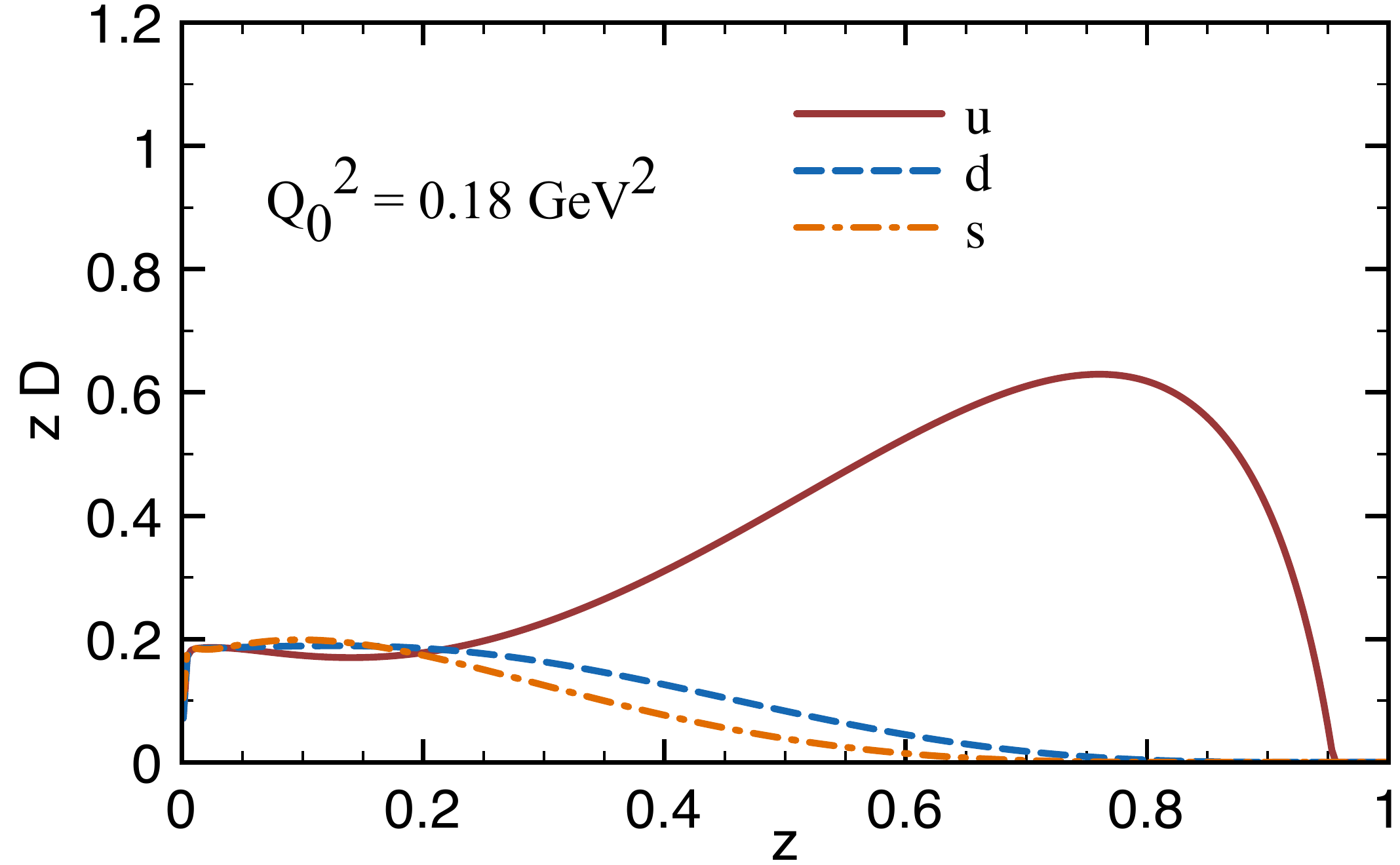}}
\hspace{0.1cm} 
\subfigure[] {
\includegraphics[width=0.4\textwidth]{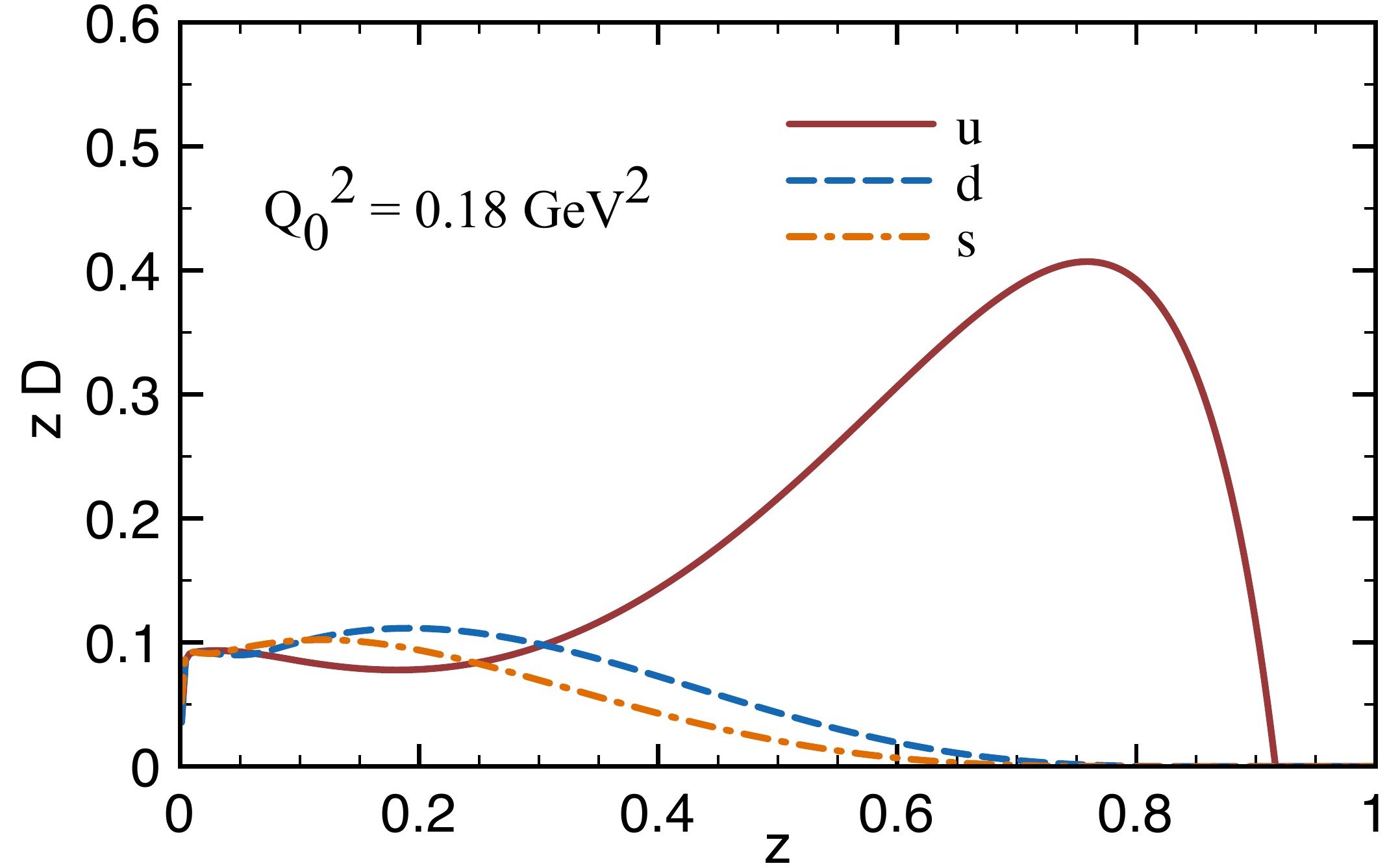}
}
\caption{a) $\pi^{+}$ and b) $K^{+}$ fragmentation functions at model scale $Q_{0}^{2}=0.18~{\rm GeV}^{2}$.}
\label{PLOT_PK_FRAG}
\end{figure}

The results for the fragmentation functions of $u$, $d$ and $s$ quarks to $\pi^{+}$ and $K^{+}$ at the model scale are shown in Fig. \ref{PLOT_PK_FRAG}. Though  the fragmentation functions of the unfavored strange and light quarks are of the same order of magnitude, we can see a notable difference between them, even at the low scale of the model. Thus one introduces considerable and ultimately unnecessary uncertainties by simply assuming that $D_{d}^{\pi^{+}}=D_{s}^{\pi^{+}}$, etc. (for example as is done in Ref. \cite{Miyama:1995bd}) .

\begin{figure}[ptb]
\setcaptype{figure} 
\centering 
\subfigure[] {
\includegraphics[width=0.4\textwidth]{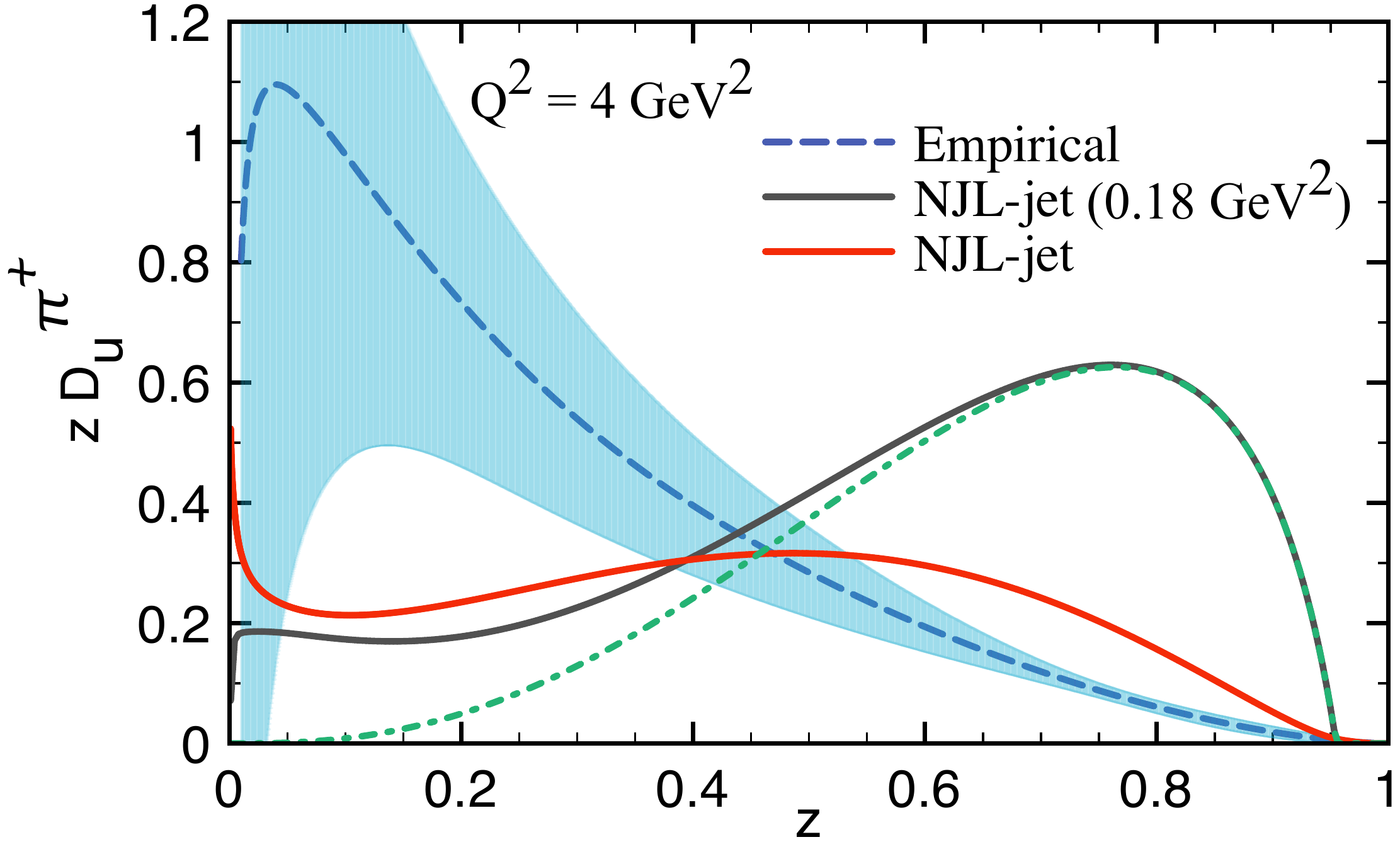}}
\hspace{0.1cm} 
\subfigure[] {
\includegraphics[width=0.4\textwidth]{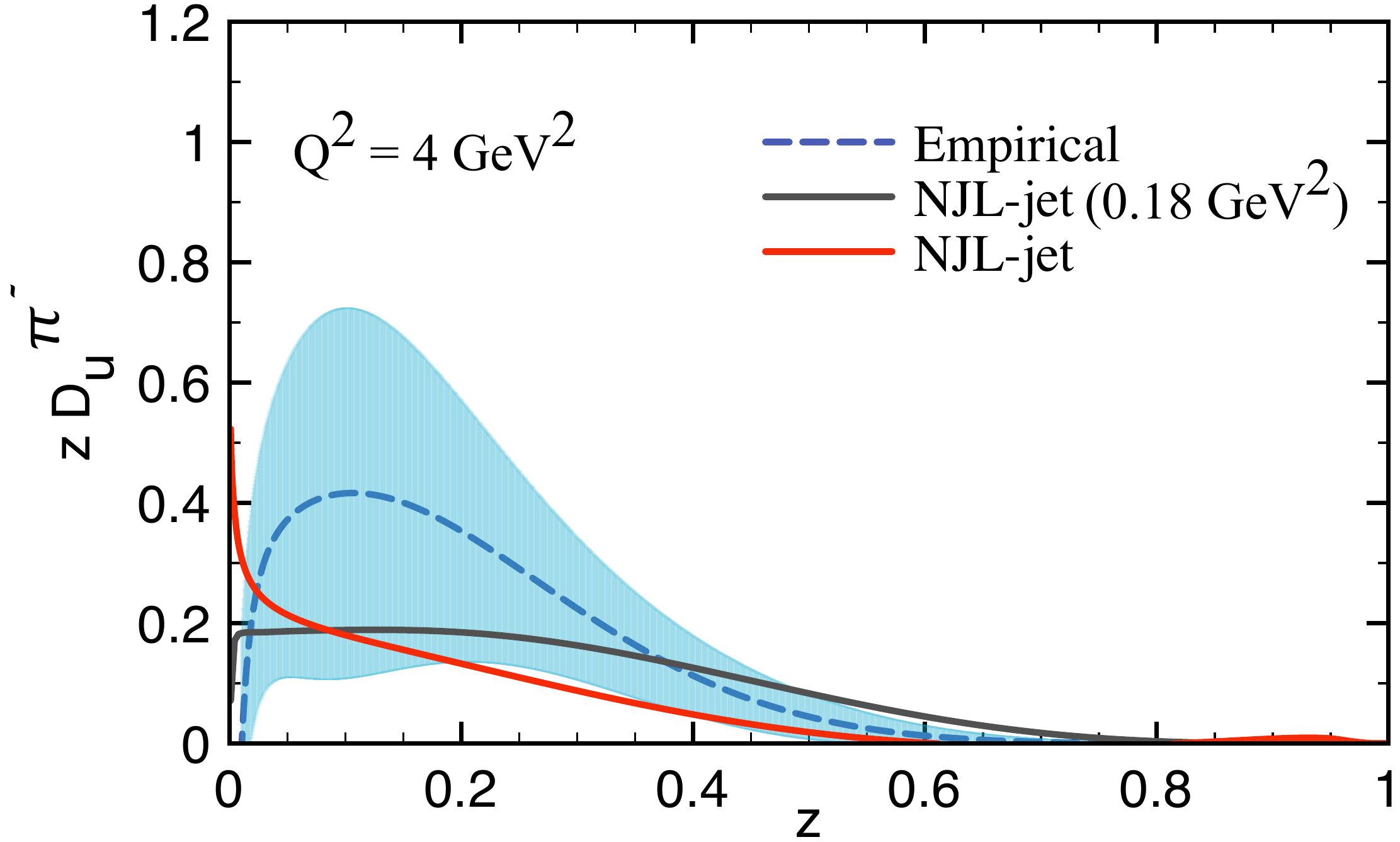}
}
\caption{Pion fragmentation functions.}
\label{PLOT_PION_FRAG}
\end{figure}

\begin{figure}[ptb]
\setcaptype{figure} 
\centering 
\subfigure[] {
\includegraphics[width=0.4\textwidth]{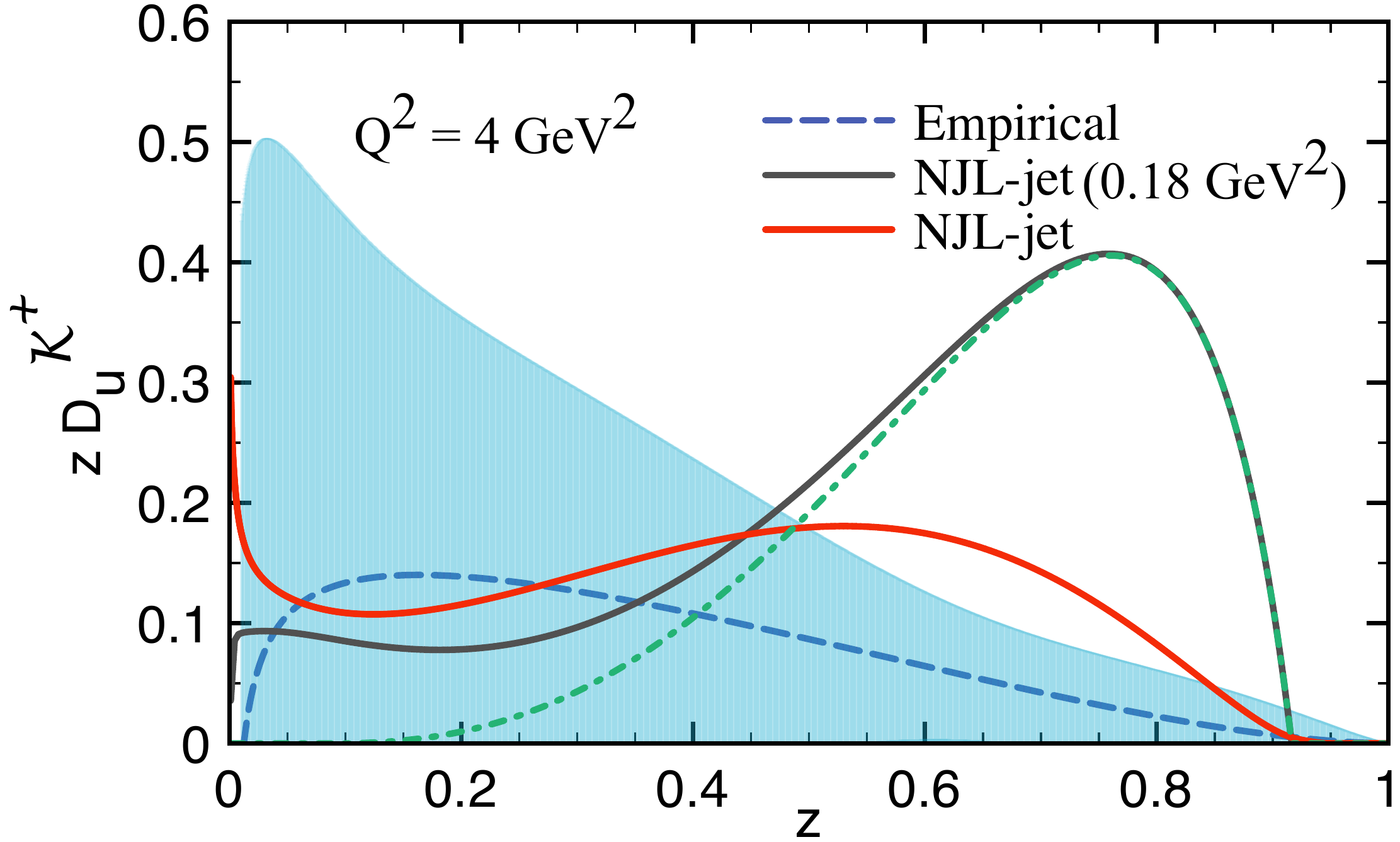}}
\hspace{0.1cm} 
\subfigure[] {
\includegraphics[width=0.4\textwidth]{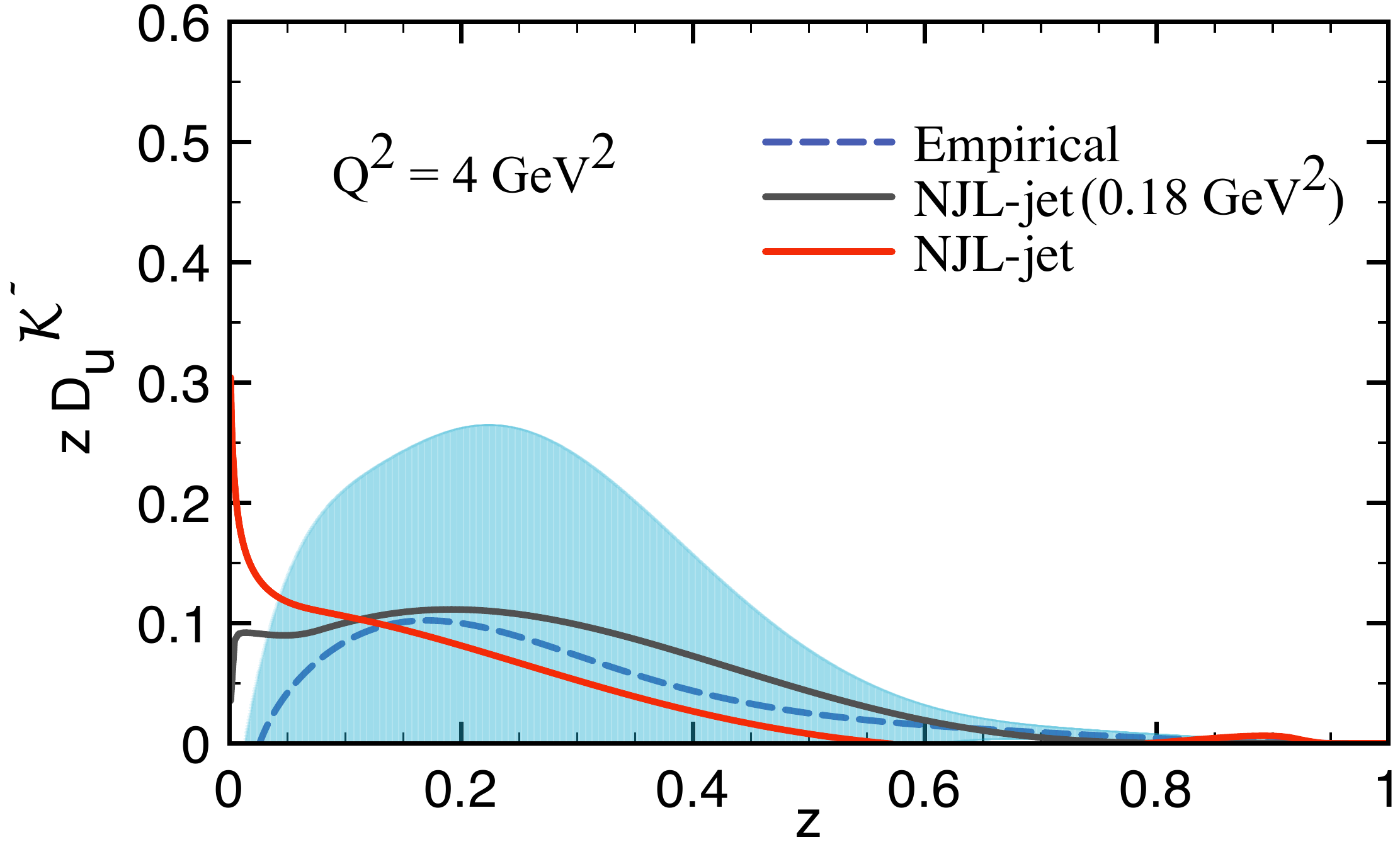}
}
\caption{Kaon fragmentation functions.}
\label{PLOT_KAON_FRAG}
\end{figure}

Further we present our results for the fragmentation functions $D_{u}^{\pi^{+}}$ and $D_{u}^{\pi^{-}}\equiv D_{\bar{u}}^{\pi^{+}}$ in Fig. \ref{PLOT_PION_FRAG}. Here the DGLAP evolved curve is also compared to the empirical parametrizations of the experimental data of Ref. \cite{Hirai:2007cx}, evolved to the same scale. We can see that the inclusion of strangeness softens the high $z$ region of $D_{u}^{\pi^{+}}$ compared to previous calculations \cite{Ito:2009zc}, thus bringing the curves closer to the phenomenological parametrizations. This is expected as the elementary fragmentation to a Kaon is not negligible. In fact the fragmentation to $K^{+}$ is about half as likely as that to $\pi^{+}$ as can be seen from the plots in Fig. \ref{PLOT_KAON_FRAG}. The plots show a reasonably good agreement with the parametrizations within the large uncertainties of the latter. 

 It is clear that for a more complete description of the quark fragmentation both vector meson and nucleon anti-nucleon channels need to be included in the calculations. This can be accomplished within the current framework.  The high $z$ region of fragmentation functions are dominated by ``few-step'' transitions where the availability of the additional fragmentation channels might have a noticeable effect. 
 
  Another limitation of the model is the assumption of the momentum scaling of the probability of hadron creation in each step of the decay chain, which is clearly only the case in the Bjorken limit. A quark with a finite momentum loses energy with each production of a hadron and finally recombines with the remnants of the antiquark jet to form the final hadron. A more accurate description of the process requires Monte-Carlo (MC) simulations of the quark fragmentation, similar to the studies in Refs. \cite{Field:1976ve, Ritter:1979mk}, and others. MC simulations would also allow one to access the transverse momentum distribution of the produced hadrons, thus becoming relevant for the analysis of a large variety of semi-inclusive data.

\section{Acknowledgements}
 
This work was supported by the Australian Research Council through the grant of an Australian Laureate Fellowship to A.W. Thomas.
  
\bibliographystyle{apsrev}
\bibliography{fragment}

\end{document}